# Orbital Kondo effect in carbon nanotubes

Pablo Jarillo-Herrero, Jing Kong[*], Herre S.J. van der Zant, Cees Dekker, Leo P. Kouwenhoven, Silvano De Franceschi[†]

*Kavli Institute of Nanoscience, Delft University of Technology, PO Box 5046, 2600 GA, Delft, The Netherlands*

[*]Present address: Department of Electrical Engineering and Computer Science, Massachusetts Institute of Technology, Cambridge, MA 02139-4307, USA

[†]Present address: Laboratorio Nazionale TASC-INFM, I-34012 Trieste, Italy

**Progress in the fabrication of nanometer-scale electronic devices is opening new opportunities to uncover the deepest aspects of the Kondo effect[1], one of the paradigmatic phenomena in the physics of strongly correlated electrons. Artificial single-impurity Kondo systems have been realized in various nanostructures, including semiconductor quantum dots[2-4], carbon nanotubes[5,6] and individual molecules[7,8]. The Kondo effect is usually regarded as a spin-related phenomenon, namely the coherent exchange of the spin between a localized state and a Fermi sea of electrons. In principle, however, the role of the spin could be replaced by other degrees of freedom, such as an orbital quantum number[9,10]. Here we demonstrate that the unique electronic structure of carbon nanotubes enables the observation of a purely orbital Kondo effect. We use a magnetic field to tune spin-polarized states into orbital degeneracy and conclude that the orbital quantum number is conserved during tunneling. When orbital and spin degeneracies are simultaneously present, we observe a strongly enhanced Kondo effect, with a multiple splitting of the Kondo resonance at finite field and predicted to obey a so-called SU(4) symmetry.**

The simplest Kondo system consists of a localized, spin-½ electron coupled to a Fermi sea via a Heisenberg-like exchange interaction[1]. This simple system can be realised with a quantum dot (QD) device[2-4], which is a small electronic island connected to metallic leads via two tunnel barriers (see Fig. 1a). Below a characteristic temperature $T_K$, the so-called Kondo temperature, a many-body singlet state is formed between the QD spin and the surrounding conduction electrons (Fig. 1a). This state adds a resonant level at the Fermi energy of the electrodes enabling the tunneling of electrons across the QD. Such a Kondo resonance can lead to a strong enhancement of the conductance overcoming the Coulomb blockade effect[2-4]. In principle, a Kondo effect may also occur in the absence of spin if another quantum number, e.g. an orbital degree of freedom, gives rise to a degeneracy (Fig. 1b). In this case, Kondo correlations lead to the screening of the local orbital "polarization", and an orbital singlet is formed through a combination of orbital states. In the presence of both spin and orbital degeneracy, quantum fluctuations lead to a strong mixing of these two degrees of freedom (Fig. 1c). This increased degeneracy yields an enhancement of $T_K$[11]. In the low-temperature limit, this system is described by a Hamiltonian obeying SU(4)-symmetry, that is, the spin and charge degrees of freedom are fully entangled and the state of the electron is represented by a 4-component "hyperspin"[12-15].

An orbital degeneracy is naturally expected in the electronic structure of carbon nanotubes[16] (CNTs). This degeneracy can intuitively be viewed to originate from the two equivalent ways electrons can circle around the graphene cylinder, that is, clockwise and anti-clockwise[17]. The rotational motion confers an orbital magnetic moment to the electrons. Consequently, the orbital degeneracy can be split by a magnetic field, $B$, parallel to the nanotube axis. (Experimental evidence for this effect, originally predicted by Ajiki and Ando[18], has been recently reported[17,19-21].) We label the orbital states of a CNT QD as $|+\rangle$ or $|-\rangle$ according to the sign of the energy shift they experience under an applied *B*. Size quantization due to the finite CNT length results in two sets of orbital levels, $E_+^{(n)}$ and $E_-^{(n)}$ where $n=1,2,3\ldots$ is the quantization number. $E_+^{(n)}=E_-^{(n)}$ at *B*=0 (assuming no orbital mixing), resulting in a four-fold degeneracy when including spin. The orbital and spin degeneracies are simultaneously lifted by a parallel *B* (Fig. 1d). The use of *B* allows tuning new degeneracies in connection with the crossing between levels from different shells. Here we are particularly interested in the crossing between states with the same spin polarisation, of the type indicated by the yellow rectangle in Fig. 1d. We show below that the two-fold orbital degeneracy originating from such a crossing gives rise to a purely orbital Kondo effect. We then consider the case of concomitant spin and orbital degeneracy at *B*=0 (green rectangle in Fig. 1d) and present evidence for an SU(4) Kondo effect.

In a measurement of the linear conductance, *G*, as a function of gate voltage, $V_G$, the four-fold shell structure leads to consecutive groups of four closely-spaced Coulomb blockade oscillations[6,22]. The *B*-evolution of such oscillations is shown in Fig. 2a for a CNT QD device (described in the inset and corresponding caption) in a $V_G$-region encompassing two adjacent shells. Coulomb peaks (highlighted by green lines) appear as lines running from bottom to top and denote the sequential addition of electrons to the QD; the electron number increasing from left to right. The observed pattern is explained in detail on the basis of the single-particle spectrum in Fig. 1d, and taking into account the Coulomb interaction between electrons (see supplementary information, SI).

The Coulomb peaks move to the left or right when increasing *B*, corresponding to adding the last electron to a $|-\rangle$ or $|+\rangle$ orbital, respectively. When the ground state configuration of

the QD changes, kinks appear in the *B*-evolution of the Coulomb peaks. The two enhanced-conductance ridges at $B=B_0\sim 6$T, bounded by two such kinks and highlighted by dotted yellow lines, are due to the crossing between $|-\rangle$ and $|+\rangle$ states as described in Fig. 1d. A detailed analysis (see SI) indicates that along these ridges the QD ground state is doubly degenerate, with the last added electron occupying the level crossing between $|+,\uparrow\rangle$ and $|-,\uparrow\rangle$ (left ridge) or between $|+,\downarrow\rangle$ and $|-,\downarrow\rangle$ (right ridge).

In the region near the degeneracy point, we are able to measure a small coupling between orbital states[21], resulting in level repulsion at $B=B_0$. The energy splitting is directly observed in the spectroscopy data of Fig. 2b where the differential conductance, *dI/dV*, is shown versus *B* and bias voltage, *V*. In this measurement $V_G$ and *B* are simultaneously varied in order to follow the middle of the Coulomb valley (dashed blue line in Fig. 2a). Here, single-electron tunneling is suppressed and the spectroscopy is based on higher-order cotunneling processes, which lead to an enhancement of *dI/dV* every time *V* equals an internal excitation energy[23]. We focus on the high-*B* region of Fig. 2b. As *B* is swept across $B_0$, the anti-crossing between $|+,\downarrow\rangle$ and $|-,\downarrow\rangle$ (depicted in Fig. 1e) shows up in the two *dI/dV* ridges highlighted by dashed yellow lines. The level spacing, corresponding to half the distance between these lines, reaches a minimum value $\delta_B=225\mu$V at $B=B_0=5.9$T. In a measurement of *dI/dV* versus $(V,V_G)$ at 5.9T, shown in Fig. 2c, the higher-order peaks appear as horizontal ridges inside the Coulomb diamond. Their spacing, $2\delta_B$, is independent of $V_G$ while their height increases towards the edges of the diamond.

An individual *dI/dV versus V* trace taken in the middle of the diamond is shown in Fig. 2d, together with traces measured at higher temperature, *T*. The strong overshoot of the *dI/dV* peaks and their log-*T* dependence (inset) indicate an important contribution from Kondo correlations. The observed behaviour is characteristic of a split Kondo resonance, i.e. a Kondo resonance associated with two quasi-degenerate states, in line with recent theoretical predictions[24] and experiments[25]. It is important to note that the Zeeman spin splitting, $E_Z=g\mu_B B_0$, is three times larger than $\delta_B$, indicating that the Kondo effect originates entirely from orbital correlations occurring at the crossing between two spin polarized states, $|+,\downarrow\rangle$ and $|-,\downarrow\rangle$. This conclusion is in agreement with the zero-field data that we show below. The large Zeeman splitting ensures also that the observed orbital Kondo resonance provides a conducting channel only for $|\downarrow\rangle$ electrons, thereby acting as a high-transmission spin filter[12-14]. On the other hand, the conductance enhancement that occurs for three-electron shell filling originates from $|+,\uparrow\rangle$ and $|-,\uparrow\rangle$ states and hence it allows only tunnelling of $|\uparrow\rangle$ electrons. Switching from one degeneracy to the other is controlled by simply switching the gate voltage, which then causes the CNT QD to operate as a bipolar spin filter.

We now centre our attention on the zero-field regime, where both orbital and spin degeneracies are expected (green rectangle in Fig. 1d). The Coulomb oscillations corresponding to the filling of a single shell are shown in Fig. 3a for a different CNT QD device. The four oscillations are clearly visible at 8K (red trace). At lower *T*, the conductance exhibits a pronounced enhancement in regions I and III, i.e., for 1 and 3 electrons on the shell, and the corresponding Coulomb blockade valleys completely disappear at 0.3K (black trace). This conductance enhancement is a hallmark of Kondo correlations. From the $T$-dependence (fully shown in the SI) we estimate $T_K$=7.7K, an unusually high value that can be ascribed to the enhanced degeneracy[11]. The important contribution of the orbital degree of freedom becomes apparent from the $B$-dependence of $G$ (Fig. 3e). If this Kondo effect was determined by spin only (this could be the case if one of the orbitals was coupled weakly to the leads), $G$ should decrease on a field scale $B \sim k_B T_K/g\mu_B$ ~6T due to Zeeman splitting[26]. In contrast, $G$ decays on a much smaller scale, $B \sim k_B T_K/2\mu_{orb}$~0.5T, which is determined by the orbital splitting (an estimate of the orbital magnetic moment, $\mu_{orb}$, is given below).

In the non-linear regime, a single zero-bias Kondo resonance appears in regions I and III (Fig. 3b). Contrary to the result in Fig. 2c, no orbital splitting is observed due to the much larger $T_K$ ($k_B T_K > \delta$[12,13,21]). In region II, we observe two peaks at finite bias, reflecting the already known splitting of a singlet-triplet Kondo resonance[27]. To show that the Kondo resonance in I and III arises from simultaneous orbital and spin Kondo correlations we investigate the effect of lifting spin and orbital degeneracies at finite $B$. As opposed to an ordinary spin-1/2 Kondo system (where the Kondo resonance splits in two peaks, separated by twice the Zeeman energy[3-9]) we find a fundamentally different splitting. At $B$=1.5T (Fig. 3c), multiple split peaks appear in regions I and III as enhanced-$dI/dV$ ridges parallel to the $V_G$-axis. In region I, the large zero-bias resonance opens up in four peaks that move linearly with $B$ and become progressively smaller (Fig. 3d). The two inner peaks are due to Zeeman splitting, i.e. to higher-order cotunneling from $|-,\uparrow\rangle$ to $|-,\downarrow\rangle$ ($|-\rangle$ is the lower-energy orbital). The two outer peaks arise from cotunneling from orbital $|-\rangle$ to orbital $|+\rangle$. In the latter case, inter-orbital cotunneling processes can occur either with or without spin flip.

(The corresponding substructure[21], however, is not resolved due to the broadening of the outer peaks.). Similar multiple splittings of the Kondo resonance have been observed also in several other samples. According to recent calculations[28], the observed multiple splitting of the Kondo resonance constitutes direct evidence of SU(4) symmetry, which implies the concomitant presence of spin as well as orbital Kondo correlations, confirming our previous finding.

The slope $|dV/dB|$ of a conductance peak (Fig. 3d) directly yields the value of the magnetic moment associated with the splitting. We obtain a spin magnetic moment $\mu_{spin} =½|dV/dB|_{spin}$ =0.06meV/T~$\mu_B$ from the inner peaks, and an orbital magnetic moment $\mu_{orb}$ =½$|dV/dB|_{orb}/\cos\varphi$ =0.8meV/T~13$\mu_B$ from the outer peaks ($\varphi$ is the angle between the nanotube and $B$)[17]. The same value of $\mu_{orb}$ follows from the splitting of the Kondo resonance in region III (Fig. 3c). In this case, however, no Zeeman splitting is observed. Here, the magnetic field induces a transition from SU(4) to a spin-based SU(2) Kondo effect for which $k_BT_K$ remains larger than the Zeeman energy, hindering the splitting of the Kondo resonance up to a few Tesla. Finally, we note that both the one-electron SU(4) and the two-electron singlet-triplet Kondo effects are characterized by a four-fold degeneracy, which results in an enhanced $T_K$[27]. Apart from this, the two phenomena are fundamentally very different. The singlet-triplet Kondo effect is a spin phenomenon in which the role of the orbital degree of freedom is simply to provide the basis for the construction of spin-singlet and triplet two-particle states (see also SI).

Since orbital Kondo correlations can only arise if the orbital quantum number is conserved during tunneling, our experimental finding of orbital Kondo physics in CNT QDs raises an interesting question concerning the nature of the dot-lead coupling. In our devices, the metal contacts are deposited on top of the CNT and the QD is formed in the segment between them[29]. It is possible that when electrons tunnel out of the QD, they enter first the nanotube section underneath the contacts, where they dwell for some time before moving into the metal. Since the orbital quantum number is likely conserved in a CNT-CNT tunnel process, this intermediate step may account for the observed orbital Kondo effect.

Correspondence and requests for materials should be adressed to P.J. (e-mail: Pablo@qt.tn.tudelft.nl)

We thank G. Zaránd, R. Aguado and J. Martinek for helpful discussions. Financial support is obtained from FOM.

The authors declare that they have no competing financial interests.

**Figure 1.** Spin, orbital and SU(4) Kondo effect in a quantum dot (QD) with an odd number of electrons. The left (right) panels in **a-c** represent initial (final) ground states. **a,** Schematic illustration of a spin-flip cotunneling process connecting the two states, spin up, $|\uparrow\rangle$, and down, $|\downarrow\rangle$, from a single orbital state. The intermediate virtual state is shown in the central diagram. This cotunneling event is one of many higher-order processes that add up coherently resulting in the screening of the local spin. **b,** Cotunneling process for spinless electrons for two degenerate orbital states, labelled $|+\rangle$ and $|-\rangle$. The depicted process flips the orbital quantum number from $|+\rangle$ to $|-\rangle$ and vice versa. The coherent superposition of orbital-flip processes leads to the screening of the local orbital quantum number. **c,** QD with two spin-degenerate orbitals leading to an overall four-fold degeneracy. Spin and/or orbital states can flip by one-step cotunneling processes, indicated by black arrows in the central diagram; the orange arrow refers to the cotunneling event connecting the two states depicted in the green diagrams. These processes lead to the entanglement of spin and orbital states resulting in an enhanced SU(4) Kondo

effect. **d,** Qualitative single-particle energy spectrum of a CNT QD in a magnetic field. Red (blue) lines represent orbital states shifting up, |+⟩, (down, |−⟩) in energy. Dashed (solid) lines represent spin up (down) states. The yellow rectangle highlights the region where a purely orbital Kondo effect can occur due to a level-crossing (at $B=B_0$) between spin-polarized states. The green rectangle highlights the SU(4) Kondo region. **e,** Zoom in on the yellow rectangle in **d**. A finite coupling, $\delta_B$, between |+⟩ and |−⟩ states causes an anticrossing (black lines). At high $B$, $\delta_B$ is smaller than the Zeeman splitting, $g\mu_B B$.

**Figure 2.** Orbital Kondo effect. **a,** Color-scale representation of the linear conductance, $G$, versus $B$ and $V_G$ at $T$~30mK ($G$ increases from dark blue to dark red). The green lines highlight the $B$-evolution of the Coulomb peaks. The dotted yellow lines highlight regions of enhanced conductance due to Kondo effect. Roman labels indicate the number of electrons on the last occupied shell near $B$=0. Orange numbers indicate the spin of the ground state. Inset: device scheme. Carbon nanotubes were grown by chemical vapour deposition on p-type Si substrates with a 250nm-thick surface oxide. Individual nanotubes were located by atomic force microscopy and contacted with Ti/Au electrodes (typical separation ~100-800nm) defined by e-beam lithography. The highly-doped Si substrate was used as a back-gate. **b,** Color-scale plot of the differential conductance, $dI/dV$, versus $V$ and $B$ along the dashed blue line in **a**. The field splits the Kondo resonance into multiple peaks. The two orange lines highlight the evolution of the peaks associated with the spin and orbital splitting, respectively. The spectroscopy features are more pronounced for $V$<0, most likely due to asymmetric tunnel barriers[30]. The yellow lines highlight the orbital anticrossing at $B=B_0$=5.9T. **c,** Coulomb diamond for 1 electron on the last occupied shell at $B$=5.9T. **d,** $dI/dV$ vs $V$

at different $T$, from 25mK (thick blue trace) to 1.1K (thick red trace), at the anticrossing point ($B$=5.9T, $V_G$=937mV). Orbital splitting, $\delta_B$, and Zeeman splitting, $E_Z$, are visually compared. The split Kondo peaks decrease with increasing $T$. Inset: peak height vs $T$ evaluated for the left peak.

**Figure 3.** Spin $\otimes$ orbital Kondo effect. **a,** Linear conductance, $G$, vs $V_G$ at $T$=8K and 0.34K. **b,** Color-scale plot of the differential conductance, $dI/dV$, versus ($V,V_G$) at $T$=0.34K and $B$=0 ($dI/dV$ increases from blue to red). **c,** Same as **b**, but at $B$=1.5T. The circle indicates the four-fold splitting in region I. **d,** Color-scale plot of $dI/dV$ versus ($V,B$) in the center of Coulomb valley I. The Kondo peak appears as a bright spot at ($V,B$) = (0,0), and splits in 4 peaks at finite $B$, following the simultaneous splitting of the orbital and spin states. **e,** $B$-dependence of $G$ taken from the zero-bias dashed yellow line in **d**. $G$ decreases on a ~0.5T scale, i.e. ~12 times faster than expected from Zeeman splitting. **f,** $G$ vs normalized Zeeman energy, $g\mu_B B/k_B T_K$ (black trace), and vs normalized orbital splitting, $2\mu_{orb}B/k_B T_K$ (blue trace). $T_K$ = 7.7 K as deduced from a fit of $G(T)$. (Note that $G(B)$=0.5$G$(0) when $2\mu_{orb}B/k_B T_K$~1). To compare the suppression due to $B$ with the one due to $V$, we also show a measurement of $dI/dV$ vs normalized bias voltage, $eV/k_B T_K$ (red trace). The blue and red traces fall almost on top of each other, implying that lifting the orbital degeneracy suppresses the Kondo effect. This demonstrates that the simultaneous degeneracy of orbital and spin states forms the origin of the strongly enhanced Kondo effect at $B$=0.

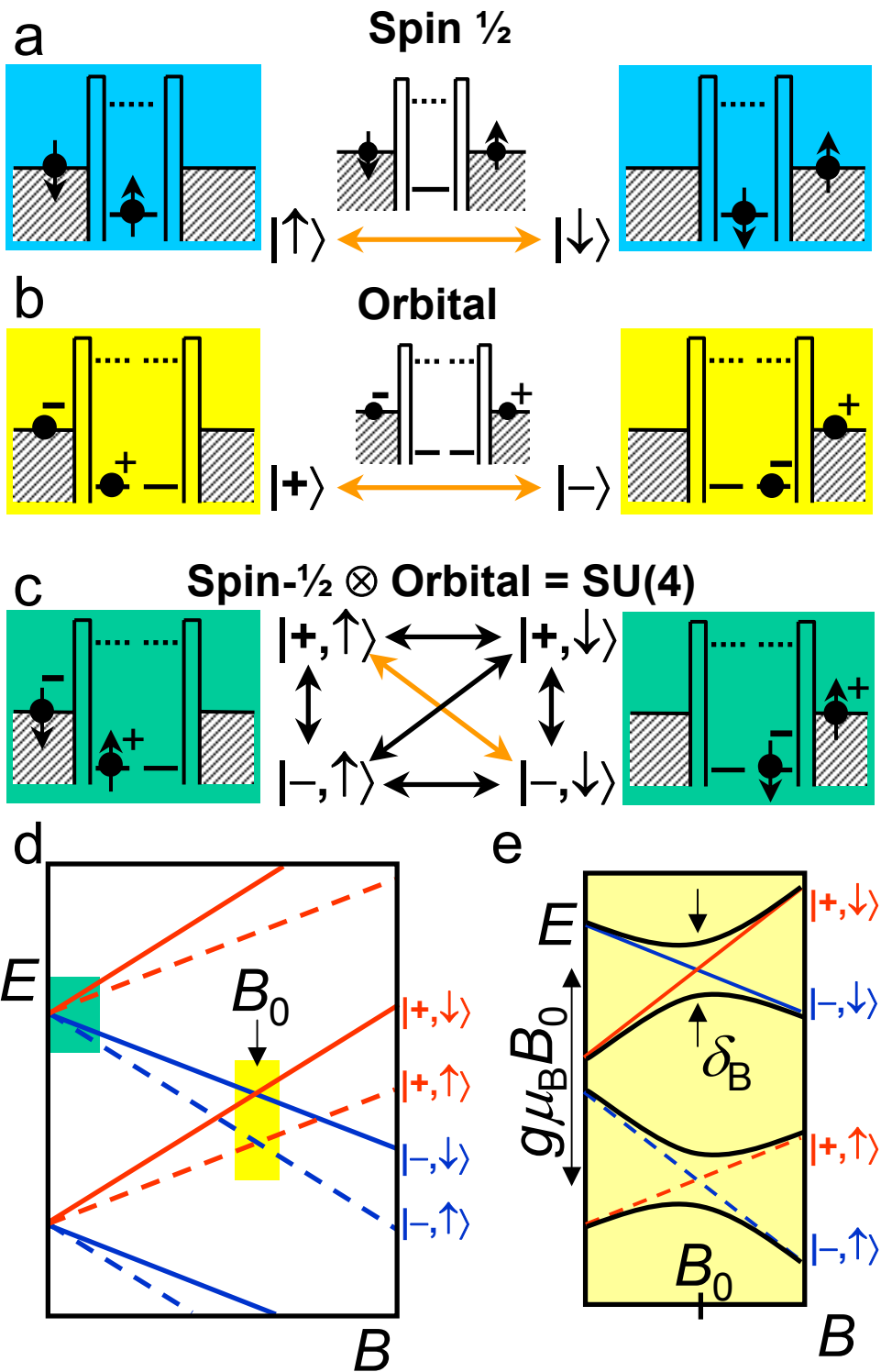

Figure 1 Jarillo-Herrero *et al.* 2004-11-26795

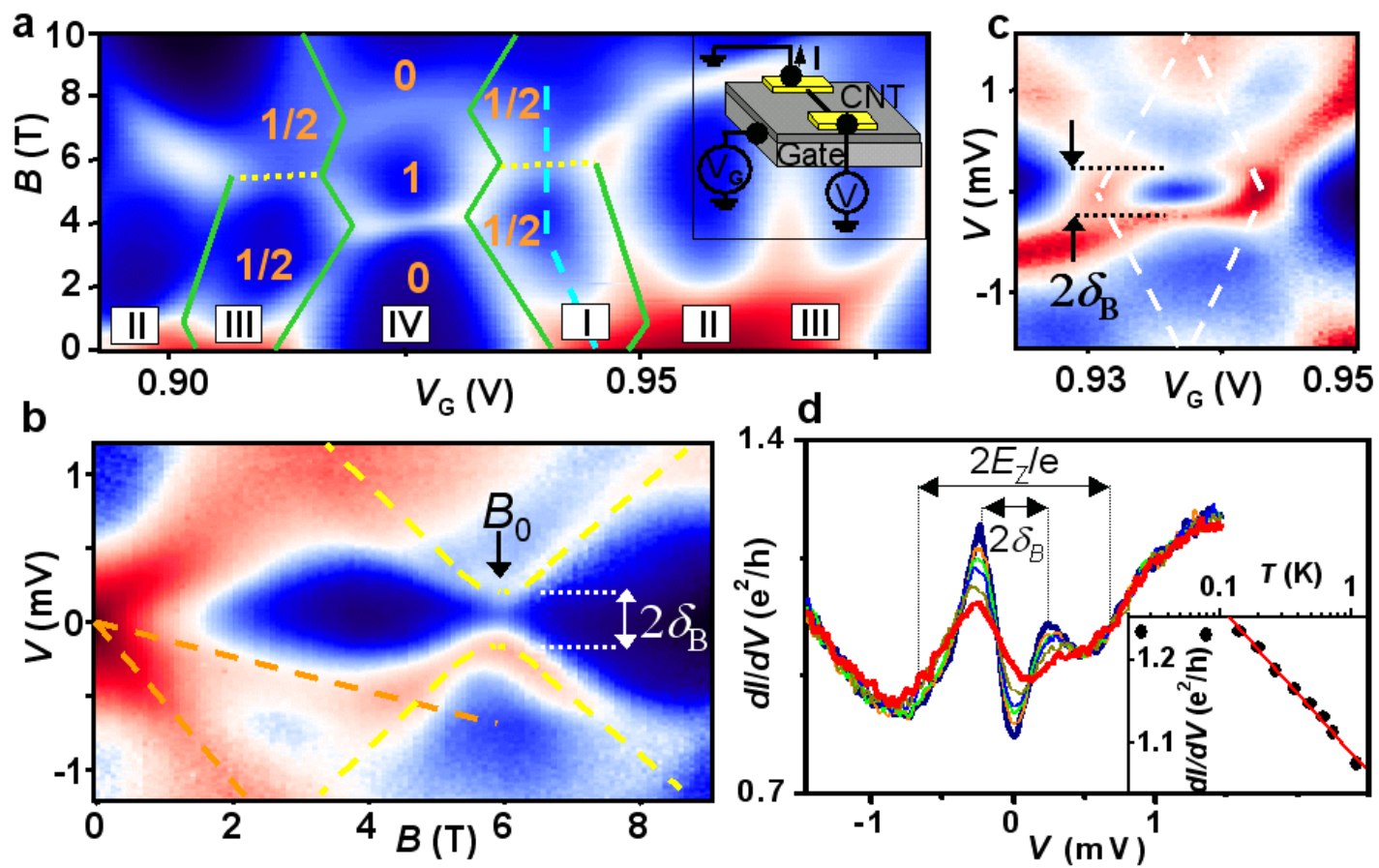

Figure 2 Jarillo-Herrero et al 2004-11-26795

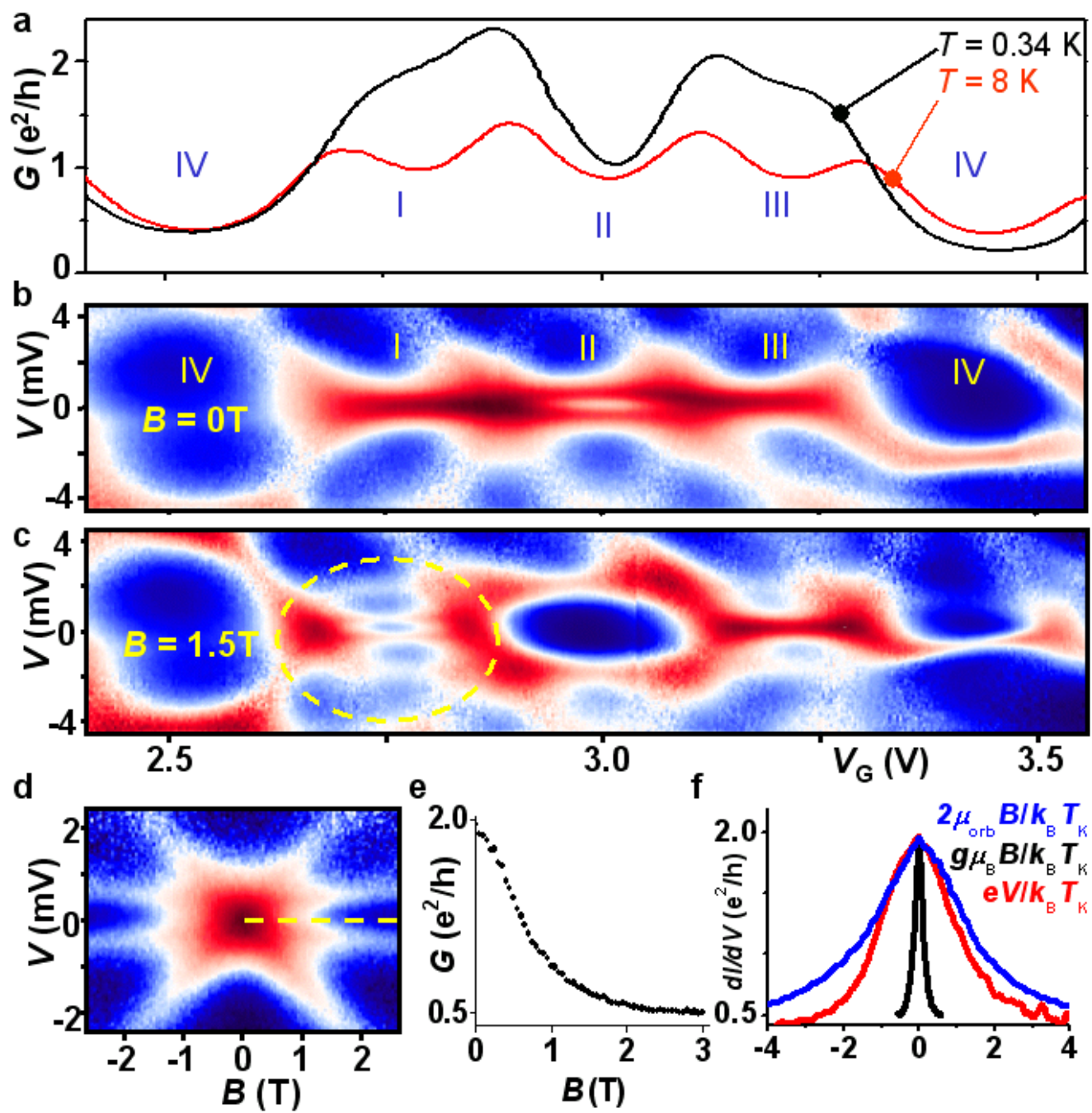

Figure 3 Jarillo-Herrero *et al* 2004-11-26795

# Supplementary Information

## Orbital Kondo effect in carbon nanotubes

Pablo Jarillo-Herrero, Jing Kong, Herre S.J. van der Zant, Cees Dekker, Leo P. Kouwenhoven, Silvano De Franceschi

### Orbital degeneracy & Kondo effect

In this section we explain the similarities and differences between the following types of Kondo effect that arise in the presence of an orbital degeneracy: two-level spin Kondo effect (TLS-Kondo), orbital Kondo effect (O-Kondo), SU(4) Kondo effect and singlet-triplet Kondo effect (ST-Kondo).

The presence of a degeneracy in the ground state is essential to all Kondo effects. In a quantum dot, the simplest Kondo effect occurs for the case of a single electron in a spin-degenerate (s=1/2) orbital level[1-3]. A degeneracy may also arise from degenerate orbitals. If this orbital degree of freedom is conserved during tunneling, then the orbital quantum number can behave as a spin, and one uses the term "pseudospin". The quantum fluctuations of this pseudospin can give rise to an O-Kondo[4-6] effect similar to the usual spin ½ Kondo effect.

In the main text we study the Kondo effect that arises from this orbital pseudospin for a single electron in a CNT QD. This pseudospin gives rise to SU(4)-Kondo when spin degeneracy is also present and to purely O-Kondo when spin degeneracy is removed. At zero field, spin and orbital pseudospin play an equivalent role and they entangle effectively with each other via cotunneling processes. Note that any of the four degenerate states is accessible via a first order cotunneling process from any of the other states (see Fig. 1c, main text). As a result of this strong entanglement the state of the dot can be mapped onto a four-component "hyper-spin" space where the Hamiltonian takes a highly symmetric form that transforms according to the SU(4) group[7-9]. We therefore denote the present effect as SU(4)-Kondo. The corresponding phase shift associated with the transmission of electrons across the dot is $\pi/4$. This sets an upper limit to the conductance of $G = 2e^2/h$ [$\sin^2 (1/2*\pi/2) + \sin^2 (1/2*\pi/2)$] = $2e^2/h$. It is worth noting that the same value for the maximum conductance is also obtained for the ordinary spin-½ and O-Kondo effects, but this time due to a $\pi/2$ phase shift and the corresponding symmetry is SU(2) ($G = 2e^2/h$ [$\sin^2 (1*\pi/2)$] = $2e^2/h$)[7-12].

The four-fold degeneracy in the SU(4)-Kondo effect leads to a Kondo temperature much higher than in the ordinary spin = 1/2 case[7-12]. A similar enhancement of $T_K$ occurs for the TLS-Kondo and ST-Kondo effects, where the degeneracy is also four-fold. In these cases, however, the physics is fundamentally different. Basically none of these Kondo effects would exist for spinless electrons, while the SU(4)-Kondo would be reduced to an SU(2) orbital Kondo effect in the absence of spin.

The ST-Kondo effect occurs for two electrons in the dot, due to the degeneracy between two-particle singlet and triplet states. It was discovered in semiconductor quantum dots[13,14] giving rise to extensive theoretical work [15-21]. In the ST-Kondo effect the orbital degree of freedom does not act as a pseudospin. Instead, it simply provides the extra orbital necessary to form the two-particle triplet state. In contrast to the SU(4)-Kondo effect, here the 4 states for ST-Kondo are not all mutually connected via a first-order cotunneling process. For example, the Sz = +1 triplet state, $|\uparrow,\uparrow\rangle$, cannot go to the triplet Sz = −1, $|\downarrow,\downarrow\rangle$, via a first order cotunneling process, since for this process $|\Delta S_z|$ = 2. The symmetry of the ST-Kondo effect is not SU(4) as it can be deduced from the analysis of the corresponding phase shifts[22,23]. It is worth noting that we also observe the ST-Kondo effect in our devices in the case when 2 electrons occupy a shell (this occurs at finite field, $B{\sim}0.35$ T, for region II (not shown), and at zero field for the shell on the left side of Fig. SI2). In our nanotube devices, as well as in semiconductor vertical quantum dots[13], the upper limit to the conductance is $G = 4e^2/h$[18,20]. Consistent with this expectation we observe a Kondo-enhanced conductance larger than the one channel limit of $2e^2/h$ (note the peak conductance of $3e^2/h$ for the shell on the left in Fig. SI2).

The TLS-Kondo and SU(4) Kondo effects are more difficult to distinguish experimentally at zero field. They occur both for a single electron in the dot. In both cases the upper limit to the conductance is $2e^2/h$ and there is a similar enhancement of $T_K$ due to orbital degeneracy. Recent calculations[24] show that the distinction is possible at finite $B$ because the Kondo resonance splits in two for TLS-Kondo (TLS-Kondo gives rise to no orbital Kondo resonance) while it splits in 4 for SU(4) Kondo. These results are in agreement with our experiments. Besides this, the Kondo effect at large $B$, due to the recovery of degeneracy between orbital states with equal spin polarization, proves the presence of orbital Kondo correlations in CNTs at zero field. Even in the presence of a coupling, $\delta$, between orbital states, since $\delta < k_B T_K$ (at $B=0$), the conditions for the observations of an SU(4) Kondo effect are fulfilled[7].

## Single particle energy spectrum & $G(V_G, B)$ spectroscopy

In this section we show how we identify the orbital Kondo effect shown in Fig. 2 with the degeneracy between the equally polarized orbital states $|+,\downarrow\rangle$ and $|-,\downarrow\rangle$ (neglecting the orbital coupling). Figure SI1a shows the same single particle energy spectrum as in Fig. 1d, where we have added the labels A, B, C and D. $\Delta$ is the energy spacing between consecutive shells. The diagrams in Fig. SI1b represents the orbital and spin configuration of the CNT QD with one electron in the highest energy shell (the lower energy shells are fully occupied) at $(E,B)$ positions A to D. In Fig. SI1c, the correspondence between the single particle energy spectrum and a $G(V_G,B)$ diagram (representing the measurements shown in Fig. 2a) is shown. U is the charging energy.

At zero field (A), the electron occupies a four-fold degenerate state, giving rise to an SU(4) Kondo effect. This shows up in Fig. SI1c as a conductance ridge inside the Coulomb blockade area. As the magnetic field is increased, the degeneracy is broken (B) and the electron occupies the lowest energy level, that is $|-,\uparrow\rangle$. At C, there is a level crossing between states $|-,\uparrow\rangle$ and $|+,\downarrow\rangle$. Due to the exchange interaction, the kink in the *B*-evolution of the corresponding Coulomb peak appears at a lower value of the magnetic field (C' in Fig. SI1b,c), when the single particle states have not crossed yet. This kink denotes a singlet to triplet transition in the region where the QD is occupied by four electrons (full shell). The singlet-triplet transition for a full shell occurs because one of the two electrons in energy level $|+\rangle$ (specifically $|+,\downarrow\rangle$) promotes to energy level $|-,\uparrow\rangle$ (from the next unoccupied shell) as soon as their energy difference is less than the exchange interaction. An enhanced conductance ridge is observed correspondingly (see also Fig. 2a). From C' on, the last added electron to the QD occupies the state $|+,\downarrow\rangle$, until a new level crossing occurs at D, between $|+,\downarrow\rangle$ and $|-,\downarrow\rangle$. Here the single electron can occupy any of the two orbital states, but in both cases with spin down. A purely orbital Kondo effect can then take place and a conductance ridge is seen (see Fig. SI1c and Fig. 2a). A similar Kondo effect can take place in region III. In this case, however, Kondo effect takes place between states $|+,\uparrow\rangle$ and $|-,\uparrow\rangle$. In a clean CNT, without disorder, this Kondo effect should take place at the same magnetic field value. Therefore a gate controlled bipolar low-impedance spin filter can, in principle, be realized[7]. By changing the gate voltage (from region I to III), we can change the filter polarity while the enhanced conductance due to Kondo effect ensures the low impedance.

## Temperature dependence

Here we show the $T$-dependence of the linear conductance data shown in Fig. 3a at $B=0$ (Fig. SI2). Starting from $T=8$ K (thick red trace), $G$ increases by lowering $T$ in the regions corresponding to partially filled shells and decreases for full shells. In the centre of valleys I and III, $G$ exhibits a characteristic logarithmic $T$-dependence with a saturation around $2e^2/h$ at low $T$, indicating a fully-developed Kondo effect (see Fig SI2, top inset, second from left). Similar $T$-dependences, although with different values of $G_0$ and $T_K$, are observed for the neighbouring shells. The coupling to the leads increases as $V_G$ decreases (the measurements are taken on the "valence band" of the small band gap)[25]. From fits to the formula $G=G_0/(1+(2^{1/s}-1)(T/T_K)^2)^s$ (ref. 26), with s=0.21, taken at the $V_G$ values indicated by arrows in Fig. SI2, we find $T_K = 6.5$, 7.7, and 16 K, respectively. These Kondo temperatures are an order of magnitude higher than those previously reported for nanotube QDs[27,28] and comparable to those reported for single-molecule devices[29,30]. Such high $T_K$ values, and the fact that $G$ exceeds $2e^2/h$ (the one-channel conductance limit) for two particles, are signatures of non-conventional Kondo effects (see **Orbital degeneracy & Kondo effect** above). The bottom inset in Fig. SI2 shows the normalized conductance, $G/G_0$, versus normalized temperature, $T/T_K$, for different shells and for both one and two electrons in the shell. The observed scaling reflects the universal character of the Kondo effect. The low-temperature behaviour is fully determined by a single energy scale, $T_K$, independent of the spin and orbital configuration responsible for the Kondo effect.

## Fabrication and measurement setup

Here we include a more detailed description about the sample fabrication procedure and measuring setup.

The nanotubes were grown by chemical vapor deposition (CVD)[31] on degenerately p-doped silicon wafers (heavy p-type, or n-type, doping is necessary to make the substrate conductive at low temperature so that it acts as a backgate) with 250nm thermally grown oxide. For the catalyst, 40mg of $Fe(NO_3)_3 \cdot 9H_2O$, 2mg of $MoO_2(acac)_2$ (SigmaAldrich), and 30mg of Alumina nanoparticles (Degussa Aluminum Oxide C) were mixed in 30ml of methanol and sonicated for ~1hr. The resulting liquid catalyst is deposited onto the substrate with $0.5\mu m^2$ openings in the PMMA resist and blown dry. After lift-off in acetone, the substrate with patterned catalyst is placed in a 1-inch quartz tube furnace and the CVD is carried out at 900°C with 700sccm $H_2$, 520sccm $CH_4$ for 10 min. Argon is flown during heating up and cooling down. The methane and hydrogen flows have been optimised to obtain long and clean nanotubes (~10μm) without amorphous carbon deposition. The nanotubes are located with atomic force microscope (AFM) inspection and e-beam lithography is then carried out to pattern electrodes over the nanotubes. Our electrodes are customized for each device, and we typically choose straight and uniform sections of nanotubes in areas free of residues. The metal electrodes are deposited via e-beam evaporation and a typical thickness of 20nm Ti and 40nm Au is used. Normally a thin Ti layer (~5nm) is used as a sticking layer, but we have noticed that thicker Ti layers (~20nm) give, on average, lower contact resistance. Nevertheless, there is always a wide distribution in the conductance of the nanotube devices, even if grown on the same substrate, probably due to the different diameters and chiralities present.

We used two setups to measure the NT-devices:

- Dilution refrigerator (Leiden Cryogenics, MK126 700) with a base temperature of 25mK (data in Fig. 2). The measurement wires have room temperature π-filters as well as low-temperature copper-powder filters. From inelastic cotunneling measurements[32,33] performed in our nanotube devices, we estimate an upper limit to the effective electron temperature of 100mK. The actual value is most likely lower, since effective electron temperatures below 50mK have been routinely observed in the same setup.

- $^3$He-system (Heliox, Oxford instruments), equipped with room temperature π-filters. Here the effective electron temperature was measured to be 340mK from fits to the width of thermally broadened Coulomb peaks in similar nanotube devices. The data in Fig. 3 and SI2 were taken in this setup.

For both setups, the differential conductance is measured with a lock-in amplifier (5μV rms excitation voltage at 17.7Hz) as a function of source-drain bias voltage and/or gate voltage. In the *T*-dependence measurements, the conductance traces are taken after the temperature has reached the desired value and stabilised for a long time (~30 min), in order to ensure proper thermalization of the sample. All temperatures quoted are cryostat temperatures.

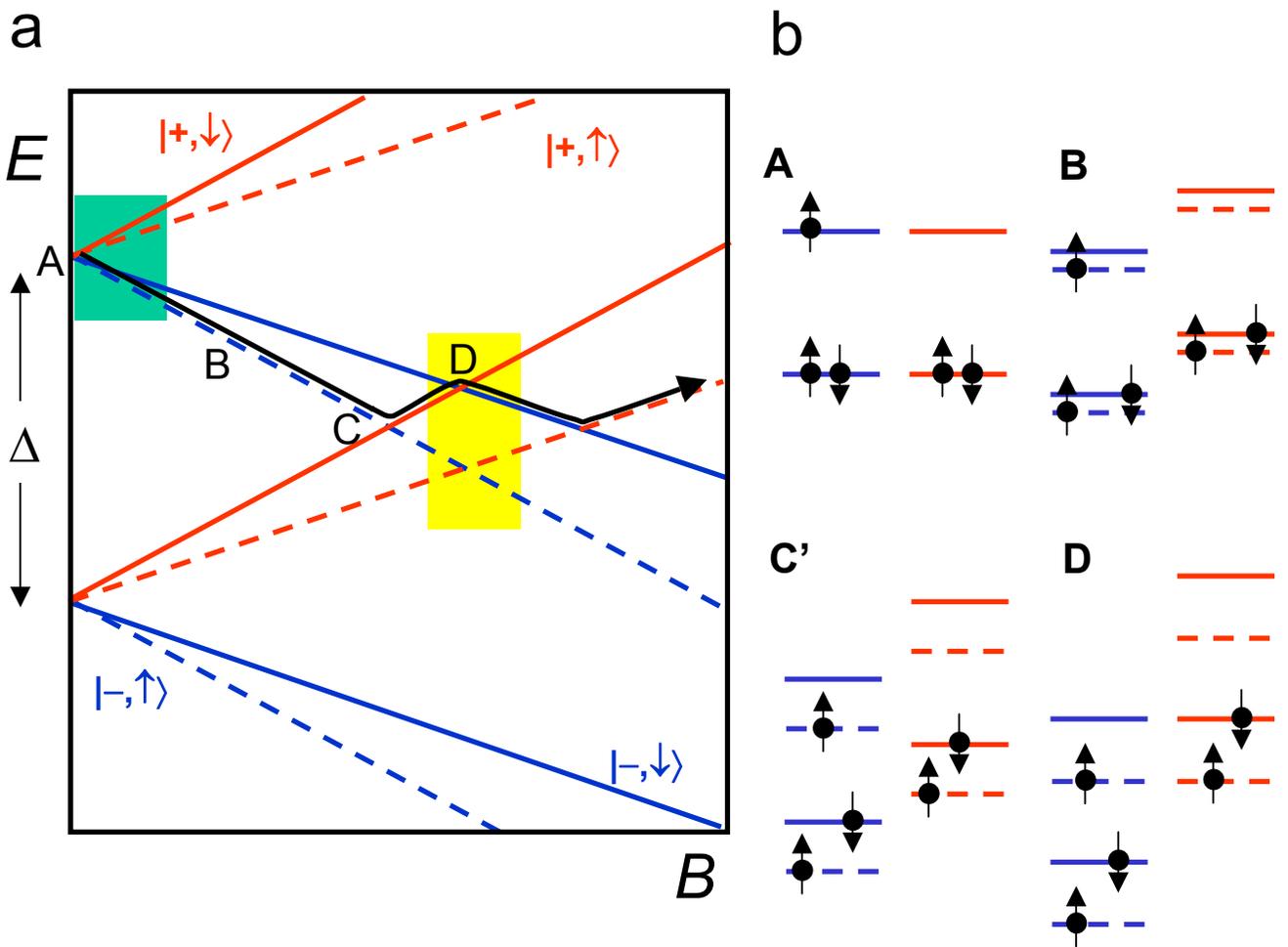

Figure SI1 Jarillo-Herrero *et al.* 2004-11-26795

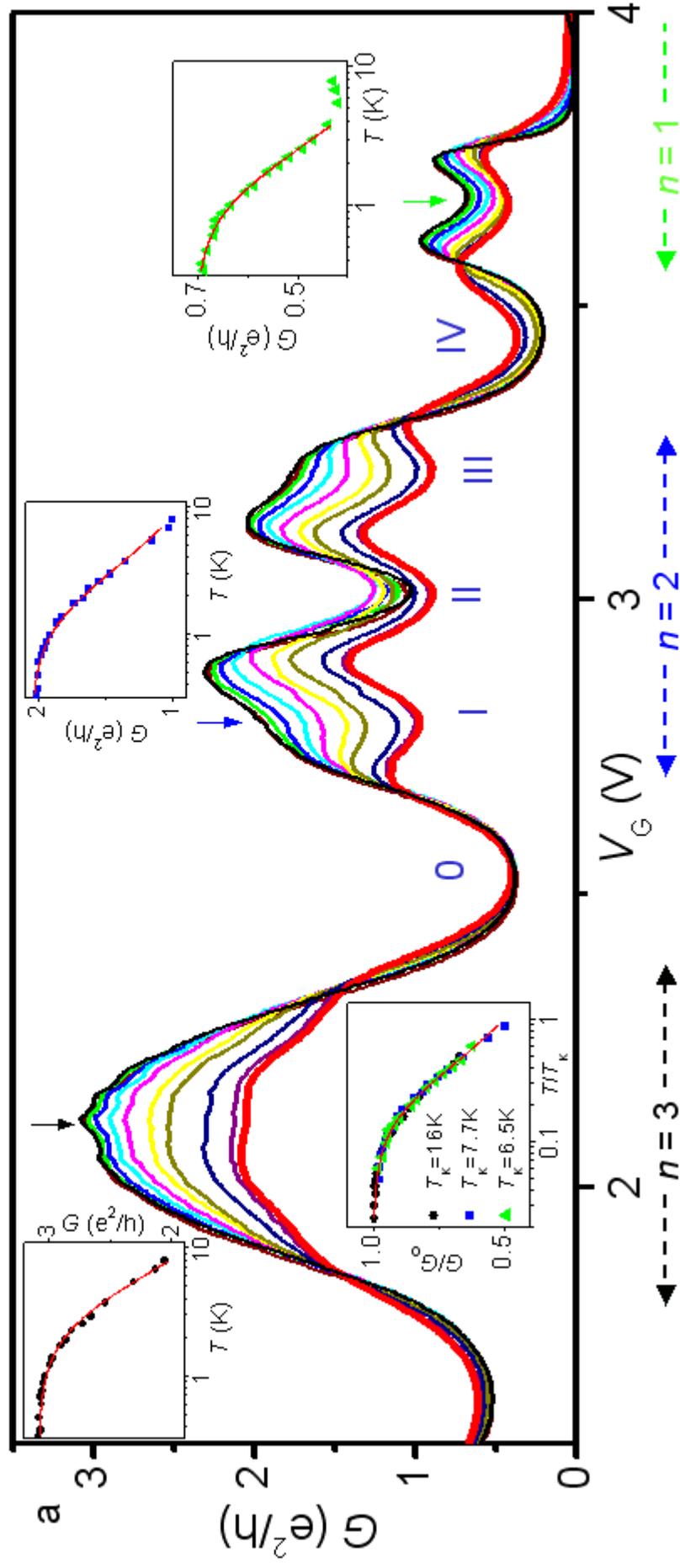

Figure SI2 Jarillo-Herrero et al 2004-11-26795